\documentclass[prb,twocolumn,amsmath,amssymb,floatfix]{revtex4-1}
\usepackage{graphicx}
\usepackage{dcolumn}
\newcolumntype{d}[1]{D{.}{.}{#1}}
\usepackage{longtable}

\usepackage{epstopdf}

\begin{document}

\title{Mott-to-Goodenough insulator-insulator transition in LiVO$_2$}

\author{Alaska Subedi} 

\affiliation{Centre de Physique Theorique, Ecole Polytechnique, CNRS,
  Universit\'e Paris-Saclay, F-91128 Palaiseau, France}
\affiliation{Coll\`ege de France, 11 place Marcelin Berthelot, 75005
  Paris, France}

\date{\today}

\begin{abstract}

I critically examine Goodenough's explanation for the experimentally
observed phase transition in LiVO$_2$ using microscopic calculations
based on density functional and dynamical mean field theories. The
high-temperature rhombohedral phase exhibits both magnetic and
dynamical instabilities. Allowing a magnetic solution for the
rhombohedral structure does not open an insulating gap, and an
explicit treatment of the on-site Coulomb $U$ interaction is needed to
stabilize an insulating rhombohedral phase. The non-spin-polarized
phonon dispersions of the rhombohedral phase show two unstable phonon
modes at the wave vector $(\frac{1}{3},-\frac{1}{3},0)$ that
corresponds to the experimentally observed trimer forming
instability. A full relaxation of the supercell corresponding to this
instability yields a nonmagnetic state containing V$_3$ trimers. These
results are consistent with Goodenough's suggestion that the
high-temperature phase is in the localized-electron regime and the
transition to the low-temperature phase in the itinerant-electron
regime is driven by V-V covalency.

\end{abstract}


\maketitle

\section{Introduction}

LiVO$_2$ occurs in a rhombohedral structure with space group
$R\overline{3}m$ above 500 K.\cite{rudo54} In this phase, the Li$^+$
and V$^{3+}$ cations and O$^{2-}$ anions occupy the sites of ordered
rocksalt structure, with the -Li-O-V-O- pattern repeating along the
rhombohedral [111] direction. This crystallographic arrangement is
characterized by alternating Li/V and O layers, where the respective
ions make a lattice of equilateral triangles within each
layer. Bongers discovered that this material undergoes a first-order
structural and magnetic transition around $T_t \approx 490$
K.\cite{bong57} The high-temperature phase exhibits a Curie-Weiss
magnetic susceptibility. The low-temperature phase does not show any
magnetic order and is nonmagnetic. Both phases are insulating.

Several competing explanations have been proposed to explain the phase
transition observed in LiVO$_2$. Goodenough has argued that the
inplane V-V separation in this material is near a critical distance
$R_c$ that corresponds to the condition where the bandwidth $W$ is
approximately equal to the on-site correlation energy
$U$.\cite{good63} According to his phenomenological
theory,\cite{good60,good91} materials that have V-V distances near
$R_c$ and $W \approx U$ exhibit an electronic instability that is
manifested by the formation of V$_3$ trimers below $T_t$. In his
words, the high-temperature phase is in the localized-electron regime,
whereas the low-temperature phase is in the itinerant-electron
regime.\cite{good91} That is, in the contemporary parlance, the
material is a Mott insulator at temperatures $T > T_t$ and a band
insulator when $T < T_t$. Each V$^{3+}$ ion in LiVO$_2$ has two
electrons in two $e_g$ orbitals, which give rise to spin $S=1$ per
site in the Mott phase. In the low-temperature ``Goodenough phase'',
the formation of V$_3$ trimers results in a gap between the bonding
and antibonding molecular orbitals of the trimers. The two electrons
per site then completely fill the bonding manifold, with the material
displaying a nonmagnetic behavior in this phase.

The trimerization envisaged by Goodenough is different from a Peierls
distortion. In a Peierls distortion, kinetic energy is gained by
moving the occupied electronic states near the Fermi level downwards,
and this changes the electronic states only near the Fermi level. By
contrast, a Goodenough distortion involves the formation of covalent
bonds between neighboring metal ions, and this changes the electronic
structure of the whole $d$ manifold. Furthermore, the Goodenough
distortion is also dissimilar to a charge disproportionation because
the charge of a V ion is transferred neither to a neighboring V nor O
ion.

Pen \textit{et al.}\ and others have proposed an alternative scenario
for the phase transition in LiVO$_2$ that involves an orbital ordering
of the localized V$^{3+}$ $3d^2$
electrons.\cite{pen97a,pen97b,ezho98,yosh11} Since the electrons
remain localized in the low-temperature phase, a structural distortion
plays a minor role in their picture, and the nature of the insulating
gap does not change across the phase transition. The orbital ordering
partitions each V layer into three sublattices, and the strong
antiferromagnetic exchange between three sites in a triangle leads to
the formation of a $S = 0$ spin-singlet state. This naturally removes
the frustration of the antiferromagnetic exchange that is present in
the high-temperature phase.

Despite having been first synthesized in 1954, experimental studies on
stoichiometric single-crystal LiVO$_2$ are relatively scarce because
of the difficulty in preparing such samples.\cite{hews85} Resistivity
measurements on polycrystalline samples of this material show an
anomaly during the phase transition at
$T_t$.\cite{reut62,koba69,hews86} Early x-ray diffraction studies
showed that there is a discontinuity in volume around $T_t$. However,
no change in crystal symmetry across the transition temperature $T_t$
was found to be apparent.\cite{koba69,bong75} Furthermore, weak
superlattice peaks that correspond to an enlarged unit cell and are
suggestive of a V$_3$ trimer formation have also been
observed.\cite{onod93,taka10,gaud13} Several independent NMR
measurements have confirmed the nonmagnetic nature of the
low-temperature phase.\cite{onod91,kiku91,jinn13} Pair distribution
function analyses of synchrotron data have additionally found evidence
for two different V-V distances below $T_t$.\cite{pour12}

Tian \textit{et al.} were able to grow single crystals of LiVO$_2$ in
2004.\cite{tian04} Their structural, magnetic, and transport studies
confirm the first-order paramagnetic-to-nonmagnetic transition in this
material. Resistivity measurements show that both phases exhibit
activated behavior with small gaps of 0.14 and 0.18 eV in the low- and
high-temperature phases, respectively. The small values for the gaps
indicate either the presence of Li deficiency or that both phases are
in the verge of metallicity. In the latter case, a strongly localized
picture may not be appropriate for the two phases. Their electron
diffraction experiments also show bright superlattice reflections that
are consistent with the trimer model.

The experimental results reviewed above give a consistent picture of
an insulator-to-insulator transition from a paramagnetic to a
trimerized nonmagnetic state in LiVO$_2$ as the temperature is
lowered. However, they do not fully clarify whether the borderline
localized-itinerant behavior proposed by Goodenough or the strongly
localized description of Pen \textit{et al.}\ is more appropriate for
this material, and the debate of whether the covalency or orbital
ordering drives the phase transition in this material remains to be
settled.

In this paper, I use calculations based on density functional (DFT)
and dynamical mean field (DMFT) theories to study the microscopic
physics underlying the high- and low-temperature phases of
LiVO$_2$. The high-temperature rhombohedral phase exhibits both
magnetic and dynamical instabilities. I was able to stabilize
ferromagnetic and 120$^\circ$ inplane ordering spin configurations in
the rhombohedral structure. However, allowing for a magnetic solution
does not open a band gap. A DMFT treatment of the V $3d$ manifold
using reasonable values of the onsite Coulomb $U$ and Hund's rule
coupling $J_H$ parameters is able to stabilize a paramagnetic
insulating state in the rhombohedral structure. The phonon dispersions
of the non-spin-polarized rhombohedral phase exhibits two branches
that are unstable along extended paths in the Brillouin zone. At the
wave vector $(\frac{1}{3},-\frac{1}{3},0)$, which corresponds to the
trimer forming instability observed in the diffraction experiments,
the two modes simultaneously show large imaginary frequencies. I was
able to stabilize a trimerized supercell corresponding to this wave
vector, and it exhibits an insulating gap. These results reconcile
well with Goodenough's suggestion that the high-temperature phase is
in the localized-electron regime and the changes in V-V bonding drives
a transition to the low-temperature trimerized phase in the
itinerant-electron regime.

\section{Computational Details}

The DFT calculations presented here were obtained using the
pseudopotential-based planewave method as implemented in the Quantum
{\sc espresso} package.\cite{qe} The phonon dispersions were
calculated using density functional perturbation theory.\cite{dfpt}
The cut-offs for the basis-set and charge-density expansions were set
to 50 and 500 Ryd, respectively, and a Marzari-Vanderbilt smearing of
0.02 Ryd was used.\cite{mv} The calculations were done within the
generalized gradient approximation of Perdew, Burke and
Ernzerhof\cite{pbe} using the pseudopotentials generated by Garrity
\textit{et al.}\cite{gbrv} I used a $14 \times 14 \times 14$ grid for
Brillouin zone sampling for the rhombohedral cell and equivalent or
denser grids for the supercell calculations. The dynamical matrices
were calculated on an $8 \times 8 \times 8$ grid, and the phonon
dispersions were obtained by Fourier interpolation. For both the high-
and low-temperature phases, the results presented here were calculated
using the structures obtained from full relaxation of the lattice
parameters in the respective phases without allowing for spin
polarization. I also performed some calculations using the general
full-potential linearized augmented planewave method as implemented in
the {\sc wien2k} package.\cite{wien2k}

The DMFT calculations were performed using tools based on the {\sc
  triqs} software package developed by Parcollet and
Ferrero.\cite{triqs} The V $3d$ subspace for the DMFT treatment was
constructed from the all-electron electronic structure using the {\sc
  triqs/dfttools} package.\cite{dftlsa,dftlsb} The local Hamiltonian
of the V $3d$ subspace has large off-diagonal elements, and the use of
the continuous-time hybridization-expansion Monte Carlo solver {\sc
  triqs/cthyb} written by Seth and Krivenko ameliorated the sign
problem.\cite{hybex,cthyb} The calculations were done at an inverse
temperature of $\beta = 23$ eV$^{-1}$ ($k_B T \simeq 500$ K $ >
T_t$). The spectral functions as a function of energy were obtained by
analytically continuing the local Green's function using the maximum
entropy method as implemented in the {\sc $\Omega$maxent}
code.\cite{oment}

\section{Results and Discussions}

\subsection{High-temperature phase}
\begin{figure}
  \includegraphics[width=\columnwidth]{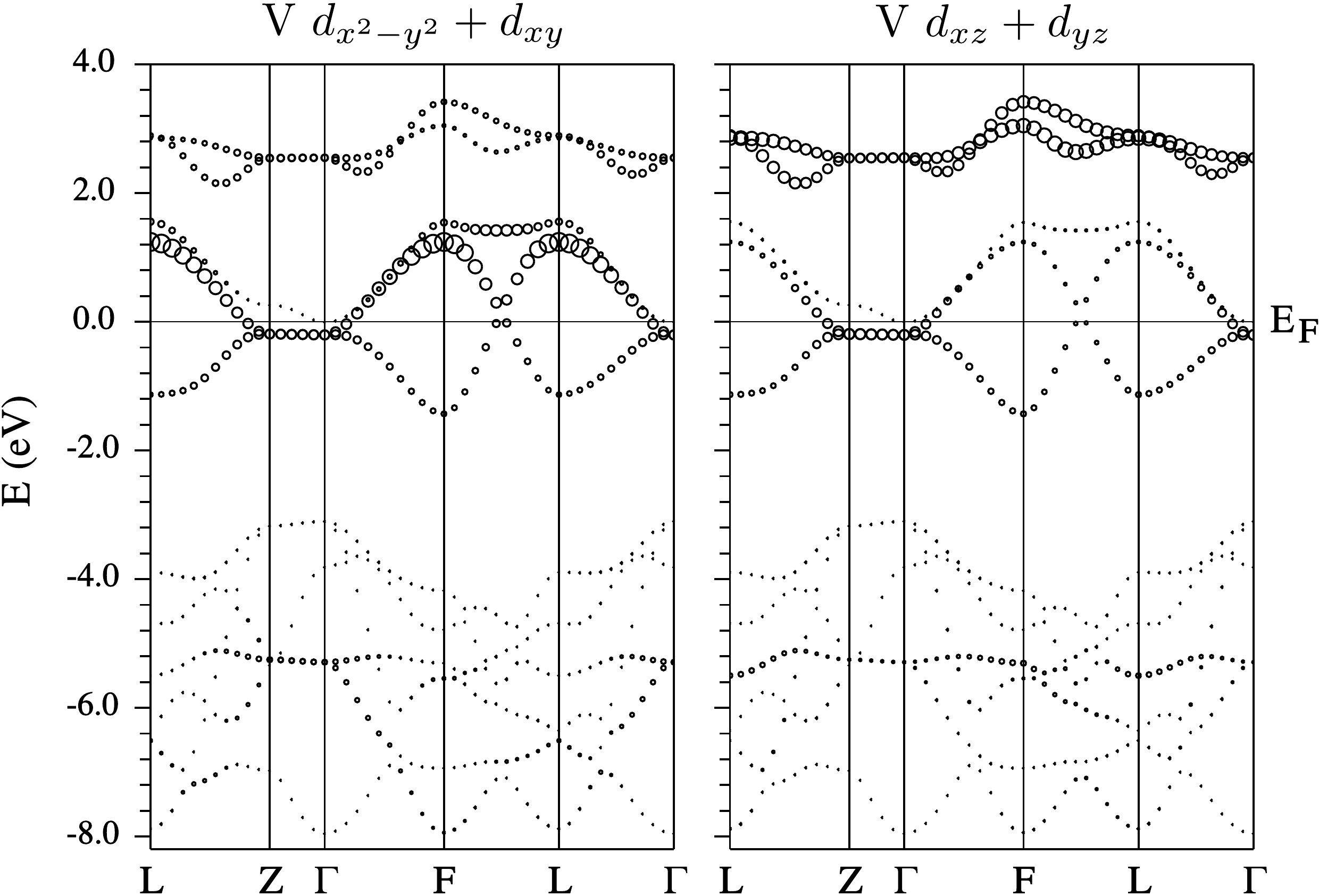}
  \caption{ Calculated non-spin-polarized band structures of
    rhombohedral LiVO$_2$ plotted with circles of size proportional to
    V $d_{x^2-y^2} + d_{xy}$ (left) and Ti $d_{xz} + d_{yz}$ (right)
    characters. The orbitals are defined in the Cartesian coordinates
    of the conventional unit cell. The band structures are plotted
    along the path $L\ (\frac{1}{2}, 0, 0) \rightarrow
    Z\ (\frac{1}{2}, \frac{1}{2}, \frac{1}{2}) \rightarrow
    \Gamma\ (0,0,0) \rightarrow F\ (\frac{1}{2}, \frac{1}{2}, 0)
    \rightarrow L\ (\frac{1}{2}, 0, 0) \rightarrow \Gamma\ (0, 0, 0)$
    in the Brillouin zone of the primitive unit cell.}
  \label{fig:eband}
\end{figure}

The calculated lattice parameters of the fully-relaxed
non-spin-polarized LiVO$_2$ in the high-temperature rhombohedral
structure are $a = b = 2.79974$ and $c = 15.03267$ \AA\ in the
hexagonal setting, which agree well with the experimental
values.\cite{rudo54} The calculated internal parameter for the O
position is $z_O = 0.24655$.  The band structure of this phase is
shown in Fig.~\ref{fig:eband}. It shows a manifold of six bands that
lie between $-$8.0 and $-$3.0 eV relative to the Fermi energy and have
predominantly O $2p$ character. The five V $3d$ bands occur between
$-$1.5 and 3.5 eV, and the Fermi level corresponds to this manifold
being occupied by two electrons.

\begin{table}
  \caption{\label{tab:occup} The orbital occupations obtained from DFT
    and DMFT calculations. The DMFT values correspond to the electron
    densities of the local Green's functions, and the values shown
    here are for the parameters $U = 5.5$ and $J = 0.8$ eV. The
    negligible asymmetry in the occupations of the degenerate orbitals
    is due to noise in the calculations.}
  \begin{ruledtabular}
     \begin{tabular}{l d{1.3} d{1.3}}
       orbital & \multicolumn{1}{c}{DFT} & \multicolumn{1}{c}{DMFT} \\
       \hline 
       $d_{z^2}$     &  0.570  & 0.099 \\
       $d_{x^2-y^2}$ &  0.428  & 0.554 \\
       $d_{xy}$      &  0.428  & 0.555 \\
       $d_{yz}$      &  0.287  & 0.394 \\
       $d_{xz}$      &  0.287  & 0.396 \\
    \end{tabular}
  \end{ruledtabular}
\end{table}

In the $\overline{3}m$ point group of the rhombohedral phase, the V
$3d$ orbitals are split into states with one nondegenerate $a_{1g}$
and two doubly degenerate $e_g$ irreducible representations. The
$d_{z^2}$ orbital belongs to the $a_{1g}$ representation, and
$(d_{x^2-y^2},d_{xy})$ and $(d_{xz},d_{yz})$ form two different basis
for the $e_g$ representation. In the electronic structure of LiVO$_2$,
the higher-lying two V $3d$ bands have a mostly $d_{xz}+d_{yz}$
character and are separated from a manifold of lower-lying three
bands. These three V $3d$ bands are derived from a pair of bands that
show a mainly $d_{x^2-y^2}+d_{xy}$ character and a band that has a
dominant $d_{z^2}$ character. As the band characters depicted in
Fig.~\ref{fig:eband} show, there is considerable mixing of orbital
characters between the two sets of $e_g$ bands. This mixing is also
evident from the orbital occupations calculated using local
projectors,\cite{dftlsa} as shown in Table \ref{tab:occup}. The
$d_{xz}$ and $d_{yz}$ orbitals that are nominally unoccupied each
contain almost 0.3 electrons, and this is due to the covalency with
the $d_{x^2-y^2}$ and $d_{xy}$ orbitals. Furthermore, I find that the
three bands in the lower V $3d$ manifold show intermixing of $d_{z^2}$
and $d_{x^2-y^2}+d_{xy}$ characters. In fact, there is substantial
$d_{z^2}$ character (not shown) in the occupied part of the nominal
$e_g$ band, which accounts for the fact that the $d_{z^2}$ orbital has
the largest occupancy with almost 0.6 electrons.

\begin{figure}
  \includegraphics[width=\columnwidth]{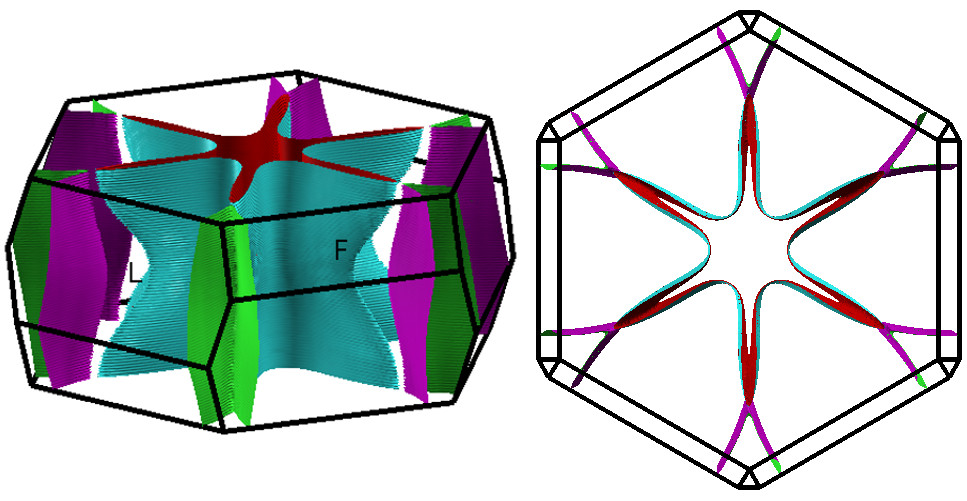}
  \caption{(Color online) Calculated non-spin-polarized Fermi surface of
    rhombohedral LiVO$_2$. The left and right panels show two
    different views of the same Fermi surface.}
  \label{fig:fs}
\end{figure}

\begin{figure}
  \includegraphics[width=\columnwidth]{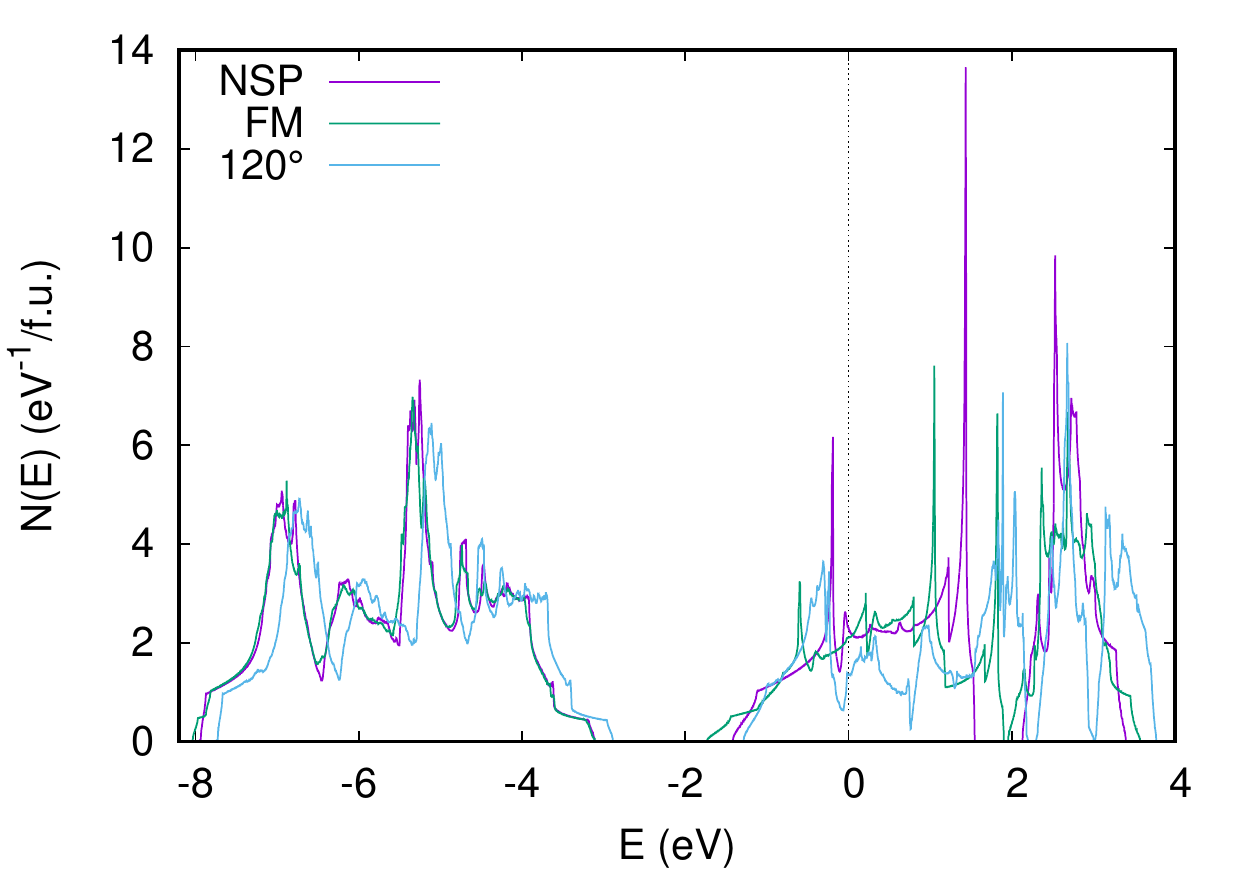}
  \caption{(Color online) Calculated DOS of rhombohedral LiVO$_2$ in
    the non-spin-polarized, ferromagnetic and 120$^\circ$ ordered
    configurations.}
  \label{fig:dos}
\end{figure}

The Fermi surface of non-spin-polarized LiVO$_2$ in the
high-temperature rhombohedral structure, shown in Fig.~\ref{fig:fs},
is rather three-dimensional. The Fermi surface does not exhibit any
obvious nesting. However, the Fermi level is situated at a place with
high density of states (DOS), as depicted in Fig.~\ref{fig:dos}. The
calculated DOS at the Fermi level is $N(E_F) = 2.3$ eV$^{-1}$ on a per
formula unit both spin basis. Using a value for the Stoner parameter
of $I = 0.8$, one gets $NI/2 > 1$, which puts this material in the
regime of ferromagnetic instability. I was able to stabilize both the
ferromagnetic and the 120$^\circ$ inplane ordering configurations of
spins. The 120$^\circ$ ordered phase is 100 meV per V lower in energy
compared to the ferromagnetic state, which itself is only 6 meV per V
lower than the non-spin-polarized state. This indicates the presence
of a strong antiferromagnetic interaction, especially considering the
relatively small values of 0.56 and 1.02 $\mu_B$ for the calculated
spin moments on the V sites in the ferromagnetic and 120$^\circ$
ordered phase, respectively. The calculated spin moments are strongly
reduced from the value of 2 $\mu_B$ expected for the $S = 1$ spins due
to localized V$^{3+}$ ($d^2$) electrons. This reduction of moments is
at odds with the results of magnetic susceptibility measurements that
show the effective moment to have a value $\mu_{\textrm{eff}} \approx
3.3$ $\mu_B$,\cite{koba69,tian04} which is close to the value of
$\mu_{\textrm{eff}}$ = 2.8 $\mu_B$ due to a localized $S = 1$
moment. This disagreement suggests that the itinerant picture of
magnetism given by DFT may not be adequate to describe the
high-temperature phase of LiVO$_2$.

\begin{figure}
  \includegraphics[width=\columnwidth]{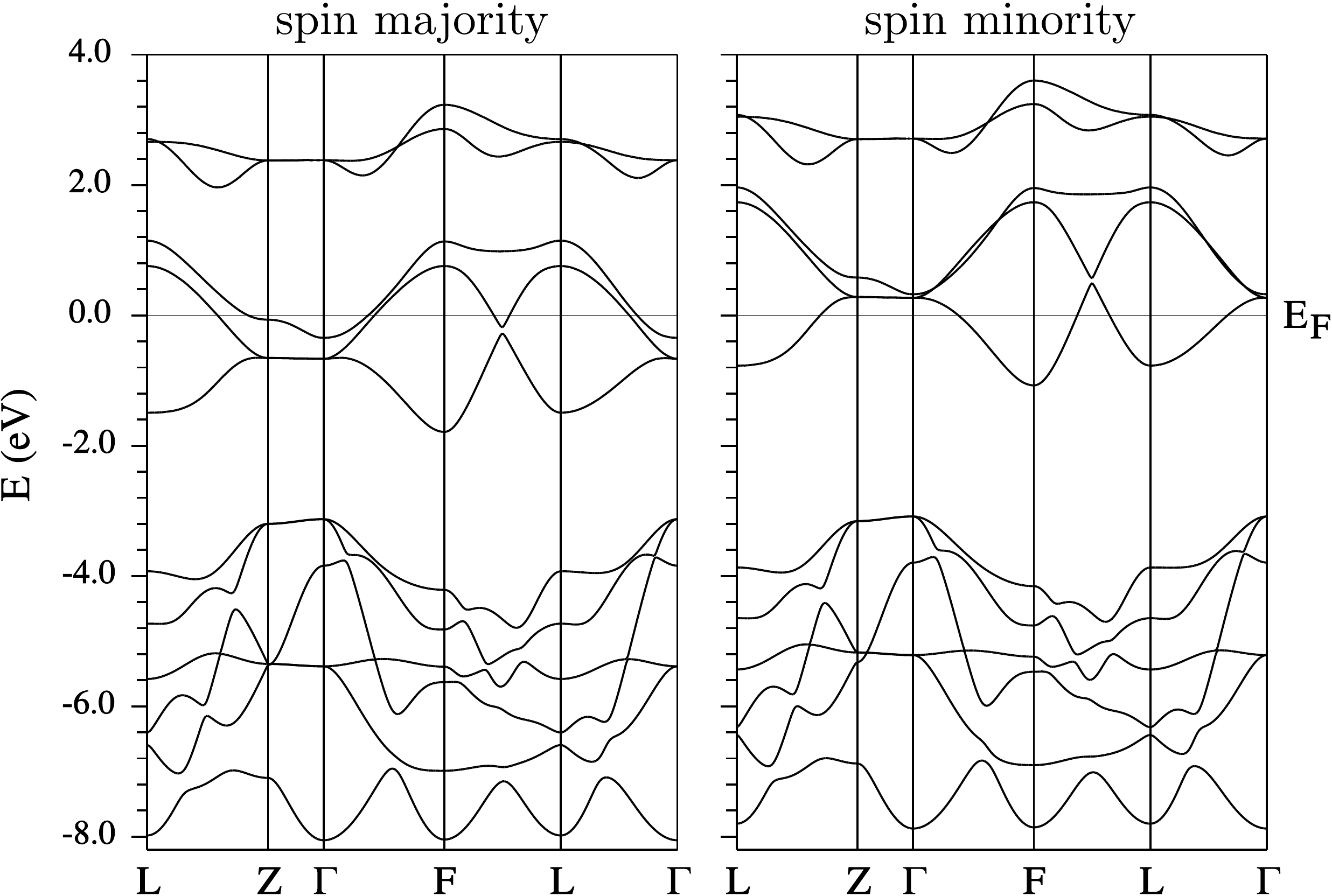}
  \caption{Calculated spin majority (left) and spin minority (right)
    band structures of rhombohedral LiVO$_2$ in the ferromagnetic
    state.}
  \label{fig:bnd-fm}
\end{figure}

The calculated DOS at the Fermi energy gets reduced when a magnetic
solution is allowed (see Fig.~\ref{fig:dos}). However, an insulating
gap does not materialize in either spin configuration, again in
contrast to the resistivity measurements that show an activated
behavior.\cite{koba69,tian04} I examined the band structure of the
ferromagnetic state, shown in Fig.~\ref{fig:bnd-fm}, to check if the
metallic state obtained from DFT calculations is due to an
underestimation of the band gap. In the spin majority sector, the
Fermi level passes through two bands. It crosses the upper part of one
nominally $e_g$ band and also goes through the lower part of the
nominally $a_g$ band. In the spin minority sector, the Fermi level
crosses the lower section of one nominally $e_g$ band. However,
increasing the exchange splitting and making the $e_g$ manifold fully
unoccupied in the spin-minority sector will not make the material
insulating because the Fermi level would still cross the partially
occupied $a_g$ band. To obtain an insulating gap, the partially
occupied $a_g$ band should be shifted above the Fermi level. Even in
the 120$^\circ$ ordered spin configuration, I find that the Fermi
level crosses through the $a_g$ band. Therefore, it seems necessary to
explicitly take into account the effects of the onsite Coulomb
repulsion $U$, which will unfill the less-than-half-filled $a_g$
band. Indeed, a DFT+$U$ study on a related material NaVO$_2$ has found
an insulating state when the $a_g$ band is completely
empty.\cite{jia09}

I performed DMFT calculations on the whole V $3d$ manifold to confirm
the above-described picture of the high-temperature insulating phase
of LiVO$_2$. It is necessary to treat all the V $3d$ orbitals within
DMFT even though the lower $e_g$ and the $a_g$ manifold are completely
separated from the upper $e_g$ manifold because the orbital characters
of the two $e_g$ manifolds are substantially mixed. A calculation that
only considers the lower three bands would in effect be neglecting the
offdiagonal parts of the local Hamiltonian and would not realistically
describe the electronic interactions present in the material.

I used $J_H = 0.8$ eV and a range of values between 4 and 6 eV for $U$
($= F^0$) for the onsite interactions in my DMFT calculations. I could
only obtain a paramagnetic insulating solution for $U > 5$ eV. The
existence of the insulating solution seems to be contingent upon a low
occupancy of the $d_{z^2}$ (\textit{i.e.}, the $a_g$) orbital. When $U
< 5$ eV, the occupancy of the $d_{z^2}$ ortibal is large (albeit
reduced from the DFT value), and the system is metallic because the
$d_{z^2}$ state shows a finite spectral weight at the Fermi level.

\begin{figure}
  \includegraphics[width=\columnwidth]{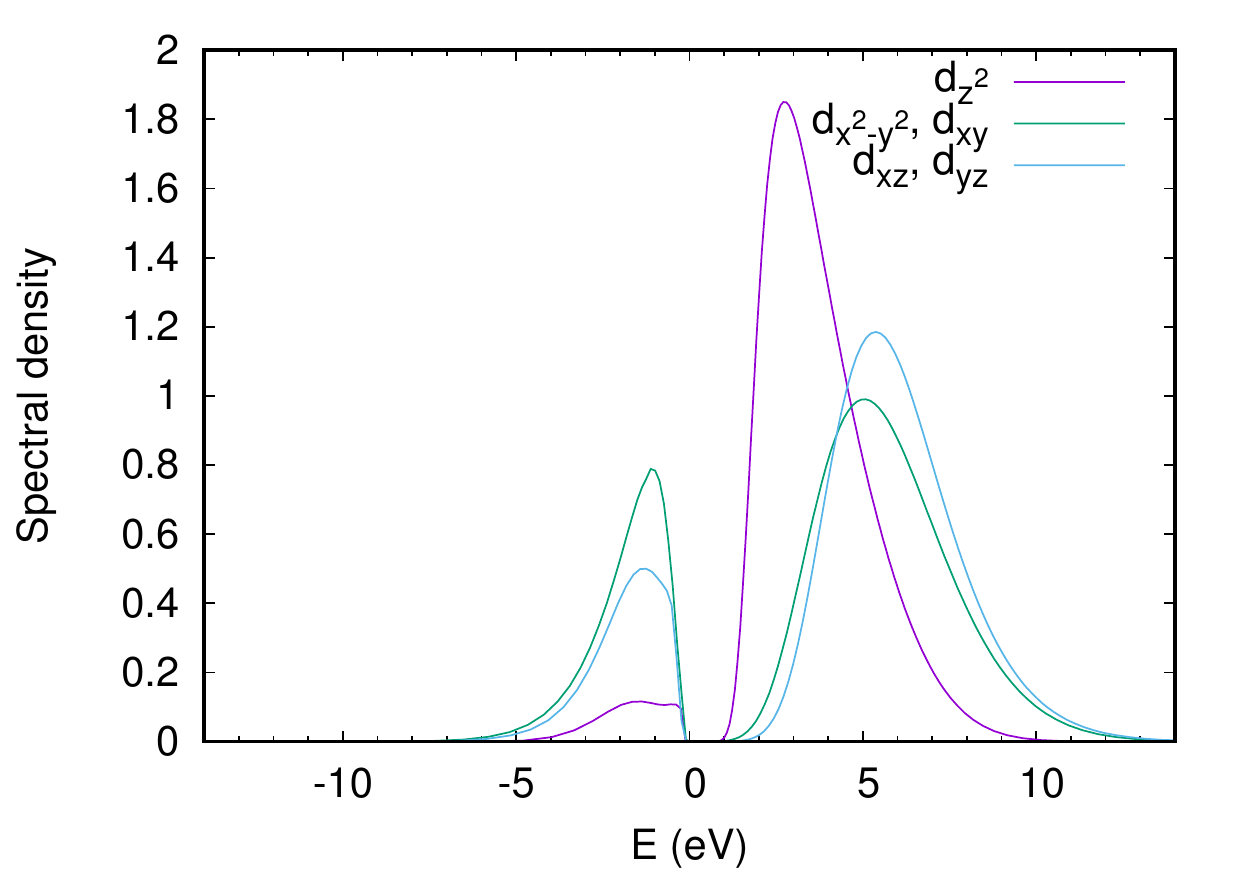}
  \caption{(Color online) Momentum-integrated spectral functions
    (local DOS) of paramagnetic LiVO$_2$ in the high-temperature
    rhombohedral structure, as obtained from DMFT calculations using
    $U = 5.5$ and $J_H = 0.8$ eV. The spectra were obtained by
    analytic continuation of the local Green's function in the
    imaginary Matsubara frequencies to real frequency using the
    maximum entropy method.}
  \label{fig:spec}
\end{figure}

The momentum integrated spectral function for the case of $U = 5.5$
and $J_H = 0.8$ eV is shown in Fig.~\ref{fig:spec} to illustrate the
electronic structure of the high-temperature paramagnetic Mott
insulating phase. The spectra display lower and upper Hubbard bands,
which correspond to the removal and addition of an electron,
respectively. The spectral weight of the $d_{z^2}$ orbital below the
gap is very small, consistent with the above discussion. The peak of
the $d_{z^2}$ spectral function lies below the peak of the upper
Hubbard band. So the insulating state is characterized by a gap
between the lower Hubbard band and the unoccupied $d_{z^2}$ band
rather than the Mott-Hubbard gap. The $d_{x^2-y^2}$ and $d_{xy}$
orbitals, which belong to the lower $e_g$ manifold, show larger
spectral weight below the gap. In addition, the $d_{xz}$ and $d_{yz}$
orbitals also show sizable weight in the lower Hubbard band. This
shows that the orbitals belonging to both the $e_g$ manifolds remain
partially occupied in the insulating state.

The paramagnetic insulating state obtained from DMFT calculations is
qualitatively consistent with Goodenough's suggestion that the
high-temperature phase of LiVO$_2$ is in the localized-electron
regime. The calculated spectral functions and the orbital occupancies
summarized in Table \ref{tab:occup} give a quantitative description of
this Mott phase, and it would be interesting to perform spectroscopy
experiments to see if these calculations realistically describe its
electronic structure.

\subsection{Low-temperature phase}

As mentioned above, LiVO$_2$ undergoes a first-order structural phase
transition from a paramagnetic insulating to a nonmagnetic insulating
phase below 490 K. The signatures of the structural transition are
seen in the x-ray and electron diffraction experiments that indicate a
tripling of the unit cell. The crystal structure of the
low-temperature phase has not been fully determined. However,
Goodenough has suggested that the low-temperature phase is
characterized by the formation of V$_3$ trimers,\cite{good63} and
recent pair distribution function analyses show the presence of two
distinct V-V distances in the low-temperature phase.\cite{pour12} It
is not known whether the positions of Li and O ions also change during
the phase transition.

\begin{figure}
  \includegraphics[width=\columnwidth]{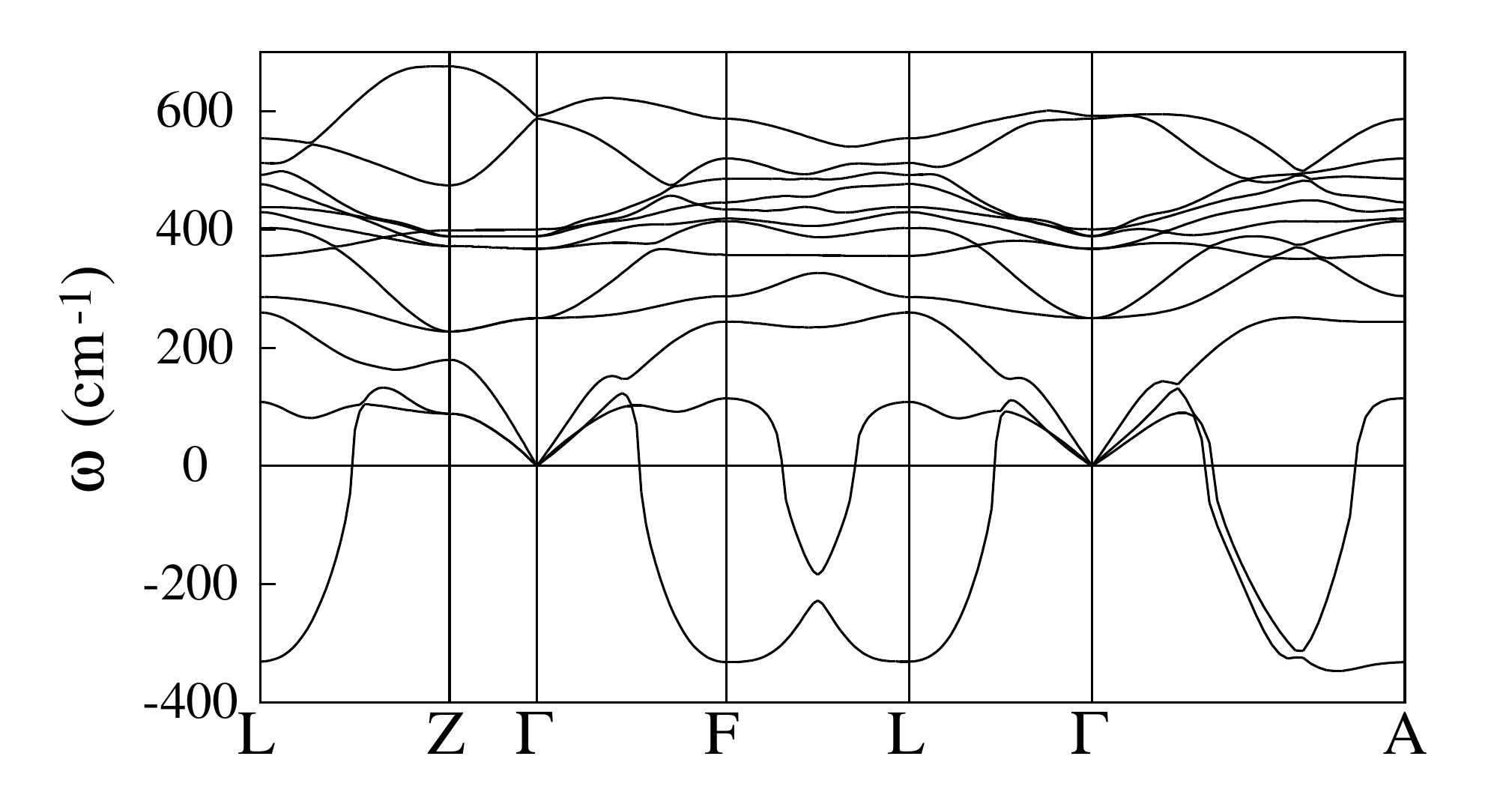}
  \caption{Calculated phonon dispersions of non-spin-polarized
    LiVO$_2$ in the high-temperature rhombohedral structure. The point
    $A$ is $(\frac{1}{2},-\frac{1}{2},0)$. The imaginary frequencies
    are represented with negative values.}
  \label{fig:pband}
\end{figure}

I calculated the phonon dispersions of the non-spin-polarized
high-temperature rhombohedral phase to investigate if it shows any
structural instabilities. As can be seen in Fig.~\ref{fig:pband}, the
phonon dispersions show two branches that are unstable along extended
paths in the Brillouin zone. These branches are connected to the two
acoustic modes near the zone center that are polarized in the inplane
direction. An inspection of the eigenvectors of these unstable modes
shows that the instability involves large inplane displacement of V
ions. Additionally, the components of these eigenvectors that
correspond to Li and O ions also show finite values, but they are
smaller than the V component. Unlike the V displacement, the Li and O
components are mostly polarized in the out-of-plane direction.

The displacement patterns of the two unstable modes differ mainly in
two aspects. First, the inplane V displacements due to these two
unstable modes are in orthogonal directions. Second, the two O ions in
the primitive unit cell move in the opposite directions along the $z$
axis for one mode, while they move in the same direction along the $z$
axis for the other mode. The eigenvectors of the two unstable modes at
the vector $(\frac{1}{3}, -\frac{1}{3}, 0)$ with respect to the
reciprocal lattice vectors of the primitive unit cell are shown in
Table \ref{tab:eigdisp}, and they exemplify the above-discussed
characteristics of the unstable modes at other points in the Brillouin
zone as well.


\newcolumntype{d}[1]{D{i}{i}{#1}}
\begin{table}
  \caption{\label{tab:eigdisp} The eigendisplacement vectors of the
    two unstable phonon modes of non-spin-polarized LiVO$_2$ in the
    rhombohedral structure at the wavevector $(\frac{1}{3},
    -\frac{1}{3}, 0)$.}
  \begin{ruledtabular}
     \begin{tabular}{l d{12.0} d{12.0} d{12.0}}
       atom & \multicolumn{1}{c}{$x$} & \multicolumn{1}{c}{$y$} &
       \multicolumn{1}{c}{$z$} \\
       \hline 
       \multicolumn{4}{c}{$\omega = 337i$ cm$^{-1}$}\\
       \hline
       Li &  0.002 + 0.000 i &  0.000 + 0.000 & -0.144 + 0.000 i \\ 
       V  & -0.799 + 0.000 i &  0.000 + 0.000 & -0.003 + 0.000 i \\ 
       O  & -0.019 + 0.000 i &  0.000 - 0.040 &  0.410 + 0.000 i \\ 
       O  & -0.019 + 0.000 i &  0.000 + 0.040 &  0.410 + 0.000 i \\ 
       \hline 
       \multicolumn{4}{c}{$\omega = 334i$ cm$^{-1}$}\\
       \hline
       Li &  0.000 + 0.000 i &  0.000 - 0.004 i &  0.000 + 0.000 i \\ 
       V  &  0.000 + 0.000 i &  0.000 - 0.807 i &  0.000 + 0.000 i \\ 
       O  &  0.017 + 0.000 i &  0.000 + 0.038 i &  0.415 + 0.000 i \\ 
       O  & -0.017 + 0.000 i &  0.000 + 0.038 i & -0.415 + 0.000 i \\
     \end{tabular}
  \end{ruledtabular}
\end{table}

The point $(\frac{1}{3}, -\frac{1}{3}, 0)$ in the Brillouin zone
corresponds to the volume tripling instability that causes the phase
transition in LiVO$_2$. Thus, the calculations of the phonon
dispersions show that low-temperature phase is stabilized by the
condensation of two unstable phonons. A symmetry analysis of the two
unstable modes shows that they belong to the irreducible
representation $A'$ of the point group $m$. Therefore, the
low-temperature structure should have the point group $m$, which has
less symmetry than the point group $\overline{3}m$ of the
high-temperature phase. Structures belonging to the $m$ crystal class
lack inversion symmetry, and the low-symmetry structure resulting from
these phonon instabilities should be polar.

I performed a full relaxation of a 36-atom unit cell to study the
structural distortions caused by the two unstable phonons identified
above. The fully-relaxed structure was refined using the {\sc spglib}
code,\cite{spglib} which resolved a primitive unit cell with three
formula units (12 atoms). The structure could also be refined to a
primitive cell with two formula units when the {\sc findsym}
code\cite{findsym} was used. The space group of the relaxed structure
is $Cm$ (no.\ 8), and this structure indeed does not have an inversion
symmetry, as expected from the discussion above.

\newcolumntype{d}[1]{D{.}{.}{#1}}
\begin{table}[b]
  \caption{\label{tab:str} Atomic positions of the fully-relaxed
    trimerized LiVO$_2$. The lattice
    parameters are $a = b = 4.86962$, $c = 5.26480$ \AA,
    $\alpha = \beta = 98.87633^\circ$, and $\gamma =
    119.94979^\circ$, and the space group is $Cm$. }
  \begin{ruledtabular}
     \begin{tabular}{l d{1.5} d{1.5} d{1.5}}
       atom & \multicolumn{1}{c}{$x$} & \multicolumn{1}{c}{$y$} &
       \multicolumn{1}{c}{$z$} \\
       \hline 
       Li &  0.01430  &  0.01430 &  0.02264 \\
       Li &  0.32383  &  0.66314 &  0.00237 \\
       Li &  0.66314  &  0.32383 &  0.00237 \\
       V  &  0.11709  &  0.82778 &  0.50433 \\
       V  &  0.54164  &  0.54164 &  0.50611 \\
       V  &  0.82778  &  0.11709 &  0.50433 \\
       O  &  0.57274  &  0.90918 &  0.74189 \\
       O  &  0.90918  &  0.57274 &  0.74189 \\
       O  &  0.24133  &  0.24133 &  0.74160 \\
       O  &  0.42252  &  0.08987 &  0.28662 \\
       O  &  0.73706  &  0.73706 &  0.23030 \\
       O  &  0.08987  &  0.42252 &  0.28662 \\
    \end{tabular}
  \end{ruledtabular}
\end{table}

\begin{figure}
  \includegraphics[width=\columnwidth]{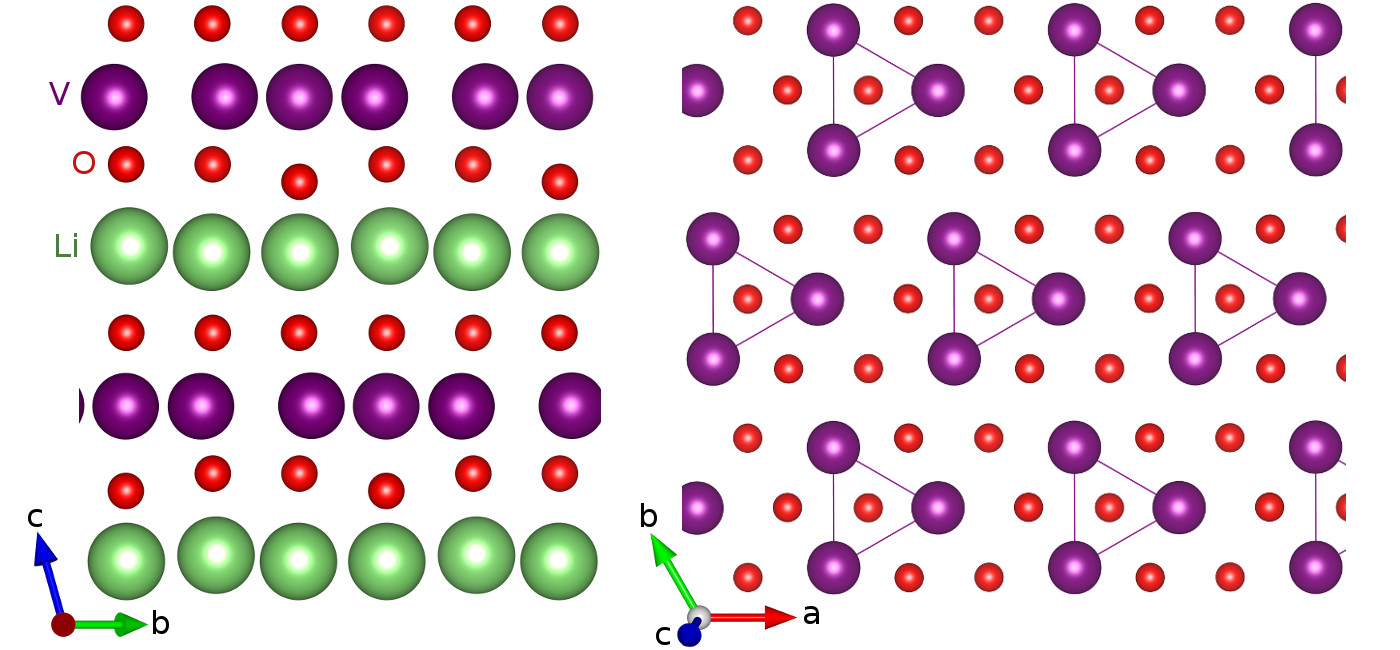}
  \caption{(Color online) The trimerized phase of LiVO$_2$ as obtained
    from structural relaxations described in the text. The left panel
    shows the layering, while the right panel shows a single VO$_2$
    layer. The V$_3$ trimers are indicated by solid lines. The large
    green, medium violet, and small red spheres denote Li, V, and O
    ions, respectively.}
  \label{fig:struct}
\end{figure}

\begin{figure}
  \includegraphics[width=\columnwidth]{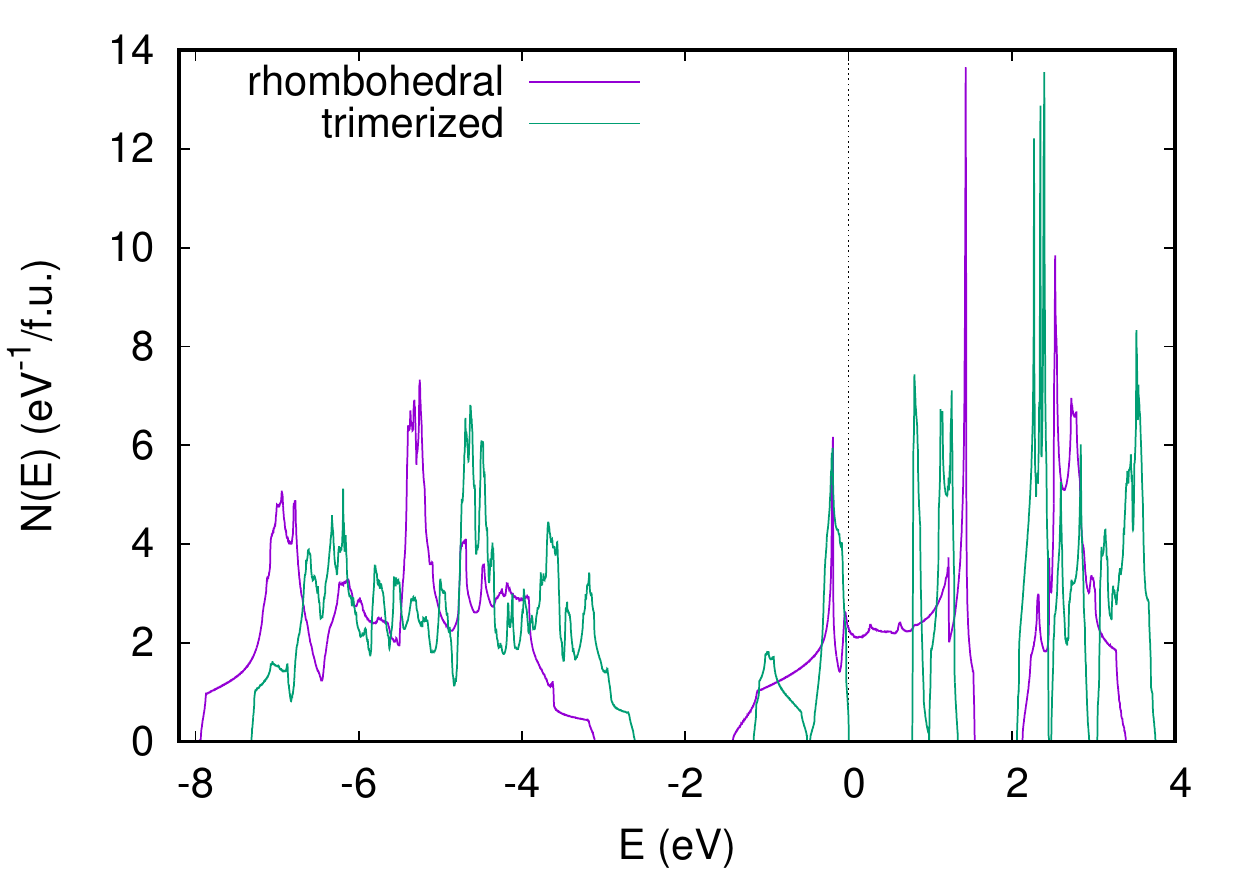}
  \caption{(Color online) Calculated non-spin-polarized DOS of
    LiVO$_2$ in the rhombohedral and trimzerized structures.}
  \label{fig:dos-tr}
\end{figure}

The full structural information of the relaxed structure is given in
Table \ref{tab:str}. This structure is graphically depicted in
Fig.~\ref{fig:struct}, and one can see that it exhibits the V$_3$
trimers predicted by Goodenough for the low-temperature phase. The
calculated DOS for this structure, shown in Fig.~\ref{fig:dos-tr},
displays a band gap of 0.78 eV, which is larger than the experimental
value of 0.14 eV obtained from transport measurements.\cite{tian04} I
did not find any magnetic instabilities in this phase. The calculated
total energy of this phase is 244 and 138 meV per formula unit lower
than the non-spin-polarized and the 120$^\circ$ ordered
high-temperature rhombohedral phases, respectively.  Therefore, this
structure likely represents the low-temperature trimerized phase
proposed by Goodenough. The DOS of both O $2p$ and V $3d$ manifolds in
the trimerized phase differ drastically from that of the
high-temperature phase. In particular, the V $3d$ manifold gets
separated into different clusters, which indicates that the V-V
bonding is different in the trimerized phase. This is in contrast to
the Peierls-type instability that only modifies the electronic
structure near the Fermi surface. The changes in the V-V bonding that
underlies the instability towards the V$_3$ trimer formation and the
resulting modification of the electronic structure over a large energy
scale is a hallmark of a Goodenough transition.

LiVO$_2$ is not the only material to show a finite wave vector
instability due to a local bonding instability. Similar mechanism has
been discussed for IrTe$_2$ and NbSe$_2$.\cite{fang13,cao13,cala09}
However, the calculated changes in the electronic structure of these
two materials are not as dramatic as in LiVO$_2$.

The microscopic calculations presented here allow me to further
explore the details of the low-temperature Goodenough phase. I find
that the V$_3$ trimers do not form equilateral triangles. Two V-V
distances within the trimer are equal to 2.43 \AA, whereas the third
V-V distance is 2.44 \AA. This distortion of the V$_3$ trimer is
small, but it is not negligible. Within the larger V$_3$ triangle, all
V-V distances are different, with values of 3.01, 3.02 and 3.03 \AA.
For comparison, I find V-V distance in the high-temperature
rhombohedral structure to be 2.80 \AA. In addition, the Li-Li
distances also change, but these changes are much smaller. There is a
smaller Li$_3$ triangle with two Li-Li distances of 2.79 \AA\ and one
Li-Li distance of 2.77 \AA, and a larger Li$_3$ triangle with two
Li-Li distances of 2.89 \AA\ and one Li-Li distance of 2.86 \AA.

Although the V-V distances change substantially as a result of the
trimer formation, I find less than 1\% difference in volume per
formula unit between the two phases, again in agreement with the
experimental observation.\cite{hews86} The VO$_6$ and LiO$_6$
octahedra keep their structural integrity and show only modest changes
in the low-temperature phase. The volume of the VO$_6$ octahedron in
the high-temperature phase is 10.87 \AA$^3$. In the low-temperature
phase, the primitive unit cell has one VO$_6$ octahedron with a volume
of 10.87 \AA$^3$ and two VO$_6$ octahedra with a volume of 10.82
\AA$^3$. Similarly, the volume of the LiO$_6$ octahedron in the
high-temperature phase is 11.81 \AA$^3$. In the low-temperature phase,
there is one LiO$_6$ octahedron with a volume of 12.08 \AA\ and two
more with a volume of 12.00 \AA.

The Li and V ions are displaced from the center of their octahedral
cage within the $ab$ plane in the low-temperature phase. But this does
not generate a large remnant electric dipole because the formation of
trimers cancels out the dipole moment on average. Nevertheless, the Li
layer and one O layer are buckled in the out-of-plane direction. [The
  other O layer is flat.] A pair each of Li and O ions are buckled in
one direction while the third Li and O ions are displaced in the
opposite direction (see Fig.~\ref{fig:struct}). Unlike the dipole
moments due to the offcenterings of Li and V ions within the $ab$
plane, the dipole moments arising from the uneven buckling of the Li
and O layers do not cancel out on average. Therefore, the
low-temperature phase of LiVO$_2$ should show a finite electric
polarization. However, it will likely be difficult to change the O-O
distances necessary to switch the polarization, and this material
cannot probably be characterized as a ferroelectric.


\section{Summary and Conclusions}

In summary, I have used microscopic calculations based on DFT and DMFT
to critically examine Goodenough's explanation for the experimentally
observed insulator-to-insulator phase transition in LiVO$_2$. I find
that the high-temperature rhombohedral phase exhibits both magnetic
and dynamical instabilities. The rhombohedral phase does not display
an insulating gap when a magnetic solution is allowed because the
spin-majority bands are always partially occupied within DFT. I was
able to obtain an insulating paramagnetic solution with reasonable
values of the onsite Coulomb $U$ and Hund's rule coupling $J_H$
parameters using DMFT calculations. The non-spin-polarized phonon
dispersions of the rhombohedral phase show large instabilities of two
modes at the wave vector $(\frac{1}{3},-\frac{1}{3},0)$, which
corresponds to the extra peaks observed in the diffraction experiments
below the phase transition. A full relaxation of the supercell
corresponding to this instability resulted in a nonmagnetic phase that
contains V$_3$ trimers. In addition, the Li and O ions are also
displaced due to this structural instability. These results support
Goodenough's suggestion that the high-temperature phase of LiVO$_2$ is
in the localized-electron regime and the V-V covalency drives the
transition to a low-temperature trimerized phase in the
itinerant-electron regime.

\acknowledgments

This work was supported by the European Research Council grants
ERC-319286 QMAC and ERC-61719 CORRELMAT and the Swiss National
Supercomputing Center (CSCS) under project s575.


\end{document}